\documentclass[twocolumn,aps,prb,superscriptaddress,nofootinbib]{revtex4-2}
\usepackage{graphicx}
\usepackage{amsmath,amssymb}
\usepackage{bm}
\usepackage{xcolor}
\usepackage{siunitx}
\usepackage{hyperref}
\usepackage{times}
\usepackage{kotex}

\hypersetup{
    colorlinks=true,      
    linkcolor=blue,       
    citecolor=blue,      
    filecolor=magenta,    
    urlcolor=blue         
}

\begin{document}

\title{Data-efficient reconstruction of critical quantum dynamics via blind fractional-envelope extrapolation}

\author{Hyunju Kim}
\email{hjkim@kentech.ac.kr}
\affiliation{Department of Energy Engineering, Korea Institute of Energy Technology, Naju-si 58330, Republic of Korea}

\author{Hyun-Yong Lee}
\email{hyunyong@korea.ac.kr}
\affiliation{Department of Applied Physics, Graduate School, Korea University,
Sejong 30019, Republic of Korea}

\author{Heung-Sik Kim}
\email{heungsikim@kentech.ac.kr}
\affiliation{Department of Energy Engineering, Korea Institute of Energy Technology, Naju-si 58330, Republic of Korea}

\begin{abstract}
Simulating real-time dynamics of quantum systems is often limited to short times by entanglement growth.
Finite-pole reconstructions such as linear prediction and related machineries extrapolate such data
reliably when the spectrum is a finite set of excitations, but at criticality the
low-energy spectrum is a power-law continuum $A(\omega)\sim|\omega|^{\alpha-1}$ --- a
branch cut whose real-time tail $G(t)\sim t^{-\alpha}$ finitely many poles cannot
represent. Here we develop a fractional-calculus-motivated envelope extrapolation for
such data. Its structure is motivated by a fractional form of Schwinger--Dyson (fSD) equation,
in which the Laplace symbol $s^{\alpha}$ carries the branch cut analytically while the
residual self-energy remains meromorphic. On real data we employ the corresponding
operational alternative --- the exponent $\alpha$ is selected blindly 
inside the fit window, the signal is detrended by $t^{\alpha}$, the residual is fitted by a
stabilized finite-pole model, and the algebraic envelope is restored. On the critical
XXZ chain this blind fractional-envelope method (fSD for short) extrapolates
short-time data typically
several-fold more accurately than finite-pole methods, with the exponent
$\alpha$ identified blindly from the fit window alone and bracketing the
closed-form Luttinger value at weak coupling. The same blind search finds the $z=2$
dilute-magnon exponent $\alpha=1/2$ at the $\Delta=1$ saturation transition, and the
advantage persists in the gapped free-magnon phase with its sharp band edges. On
noncritical dynamical mean-field spectra, whose low-frequency response is regular, fSD
by contrast fails to select any stable fractional envelope and reduces to the standard
pole result rather than manufacturing a spurious power law, making it an efficient and accurate route to quantum
critical dynamics when only short simulation times are accessible.
\end{abstract}

\maketitle

\section{Introduction}
Real-time correlation functions are the most direct bridge between microscopic models
and spectroscopy: the single-particle spectral function, the dynamic structure factor,
and transport coefficients are all encoded in the time evolution of two-point functions.
Tensor-network methods compute these correlators essentially without statistical noise in one- or two-dimensional systems
and treat low- and high-energy features on an equal footing, but they share a hard
limitation: the entanglement generated by real-time evolution grows with time, so the
bond dimension required to represent the state --- and hence the cost --- grows with the
simulated time~\cite{White2004,Daley2004,Haegeman2011}. 
Hence the low-frequency, long-time behavior that governs critical and
hydrodynamic phenomena often lies beyond the reach of a computationally accessible time window. 

Two complementary strategies address this. The first suppresses the entanglement growth by
evolving along a contour in the complex-time plane and reconstructing the real-time /
real-frequency result, a now-established and accurate route for impurity and lattice
problems~\cite{Cao2024,Grundner2024,Yu2026}. The second extends the available data by
representing the correlator as a sum of decaying complex exponentials --- equivalently, a
finite set of poles of the resolvent --- using linear prediction (LP), the matrix-pencil and ESPRIT (estimation of signal
parameters via rotational-invariance techniques) methods, the AAA (adaptive
Antoulas--Anderson) rational-approximation algorithm, or minimal-pole
estimators~\cite{WhiteAffleck2008,Roy1989,Hua1990,AAA2018,ZhangGull2024,HuangGullLin2023},
with recent work pushing exponential fitting explicitly toward long-time extrapolation and
infinite-time limits~\cite{Erpenbeck2026,Kemper2024}, including dynamic mode
decomposition (DMD) --- essentially a matrix-pencil variant --- applied to forecasting
quantum many-body dynamics~\cite{Kaneko2025}. Where the spectrum is
\emph{pole-dominated} --- a Fermi-liquid quasiparticle, a Kondo resonance, or gapped Hubbard
bands --- these methods are excellent: the analytic structure of
the resolvent is genuinely a finite collection of poles, and a finite exponential sum is
the right model.

\begin{figure*}[t]
\includegraphics[width=0.80\linewidth]{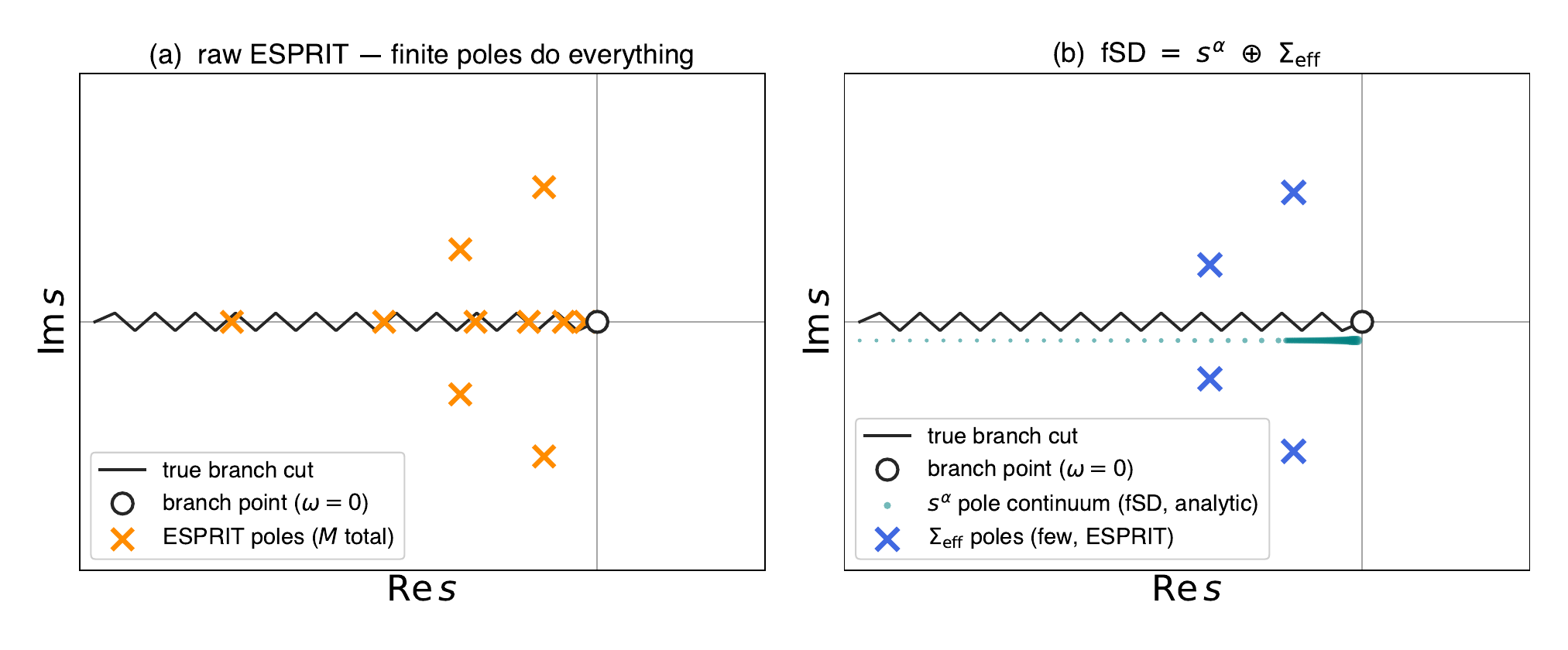}
\caption{\label{fig:schematic}
Analytic structure underlying the method, in the complex-frequency ($s$) plane. In both
panels the true critical branch cut (zigzag line) runs along the negative real axis and
terminates at the branch point $s=0$ (open circle). (a) A finite-pole estimator must
represent the dynamics with discrete poles ($\times$); to mimic the cut it tiles the
negative real axis, under-resolving the $\omega\to0$ edge and missing the asymptotic
$t^{-\alpha}$ tail. (b) The fractional closure writes the inverse propagator as
$s^{\alpha}+\tilde\Sigma_{\rm eff}(s)$: the cut is carried analytically by $s^{\alpha}$,
so that the propagator behaves as $s^{\alpha-1}$ near the branch point; the pole
continuum shown (offset for clarity), with power-law weight growing toward the branch
point, is the Stieltjes representation of that propagator singularity
[Eq.~\eqref{eq:stieltjes}], while only the meromorphic residual
$\Sigma_{\rm eff}$ (a few poles) is estimated by ESPRIT.}
\end{figure*}

On the other hand, at a quantum critical point, the low-energy
spectral function is a power-law continuum, $A(\omega)\sim|\omega|^{\alpha-1}$, 
which corresponds to an algebraic tail of the correlation function $G(t)\sim t^{-\alpha}$ rather than a sum of
exponentials. In the complex frequency plane this is a \emph{branch cut}, not a pole. 
A finite sum of poles can mimic a branch cut by tiling it with a dense line of poles, especially with geometrically distributed decay rates, but maintaining a fixed relative accuracy as the dynamic range extends toward $t\to\infty$ (hence $\omega \to 0$) requires an increasing number of modes. Rational approximants can represent a branch cut efficiently on a prescribed finite domain through poles clustered near the branch point, with root-exponential convergence for algebraic singularities. However, extracting the asymptotic branch-point exponent from a short time window remains an ill-conditioned inverse problem, because the relevant pole accumulation closest to the branch point is controlled by times beyond the observed window (Fig.~\ref{fig:schematic}) \cite{Stahl1993,Beylkin2005,Beylkin2010}.

Here we aim to encode the critical power-law memory directly, inspired by a fractional
Schwinger--Dyson (fSD) expression --- with the fractional order $\alpha$ selected from
the data --- in which $\alpha$ embeds the branch
cut analytically through the Laplace symbol $s^{\alpha}$, while the residual self-energy
$\Sigma_{\rm eff}$ remains meromorphic. Such a fractional differential equation approach has previously been applied to
encode the power-law memory of open quantum systems~\cite{PengZhang2025}. In this work, 
in combination with the previously suggested complex-time evolution technique~\cite{Cao2024,Yu2026,ChaLeeKim2025}, we demonstrate the capability of the fSD approach 
in the one-dimensional $S=1/2$ XXZ chain in the critical regime. 
Our contribution is twofold; {\it i)} the data efficiency of the fSD approach compared
to finite-pole extrapolation methods (LP, ESPRIT, and an ESPRIT-variant long-time
extrapolator by Erpenbeck {\it et al.}~\cite{Erpenbeck2026}), and {\it ii)} blind (data-selected) identification of the critical
exponent, with the blind exponent feeding the extrapolation so that the two are not
contaminated by initial bias. 
In addition, we demonstrate the method's falsifiability on the single-orbital
Hubbard model on the Bethe lattice, solved via dynamical mean-field theory
(DMFT)~\cite{Georges1996,Yu2026}, which is a pole-dominated system. We show that the data-efficiency advantage collapses to unity, and that fSD
recognizes the absence of a branch cut from the data and reduces to the standard pole
result, rather than manufacturing a spurious power law. 

\section{Theoretical background}
\label{sec:theory}
\subsection{Coherent poles versus branch cuts in correlation functions}
Throughout this work $G(t)$ denotes a generic time signal to be extrapolated; the
spectral constraints attached to it depend on the operator class. For a fermionic
single-particle Green's function, the retarded function is defined with an
anticommutator, $G^{R}_{f}(t)=-i\theta(t)\langle\{c(t),c^{\dagger}\}\rangle$, and its
Lehmann representation
\begin{equation}
G^{R}_{f}(z)=\int d\omega'\,\frac{A_{f}(\omega')}{z-\omega'},\qquad
A_{f}(\omega)=-\tfrac1\pi\,\mathrm{Im}\,G^{R}_{f}(\omega+i0^{+}),
\end{equation}
is analytic in the upper half plane, with a nonnegative spectral function obeying the
zeroth-moment sum rule $\int A_{f}\,d\omega=1$. For a bosonic or spin response 
we use the dynamic structure factor $S(\omega)=\int dt\,e^{i\omega t}\langle \hat{O}(t)\hat O^{\dagger}\rangle$ associated with $C(t)=\langle\hat O(t) \hat{O}^{\dagger}\rangle$. The XXZ calculations below extrapolate $C(t)=\langle S^{x}_{i}(t)S^{x}_{i}\rangle$, whereas the DMFT
calculations extrapolate the fermionic retarded Green's function. 

For either class the relevant spectral weight [$A_{f}(\omega)$ or $S(\omega)$; we
write $A(\omega)$ generically] can exhibit two qualitatively different analytic
structures, with qualitatively different time dependence. \emph{Poles:} a finite sum
of broadened delta functions gives $G(t)=\sum_k w_k e^{-i\omega_k t-\gamma_k t}$,
corresponding to the Fermi-liquid, Kondo, or Hubbard-band case. \emph{Branch cut:} if the spectral weight near a spectral edge $\omega_{\rm edge}$ behaves as $A(\omega_{\rm edge}+\epsilon)\sim \epsilon^{\alpha-1}$, the leading long-time contribution to its Fourier transform gives $G(t)\sim e^{-i\omega_{\rm edge}t}t^{-\alpha}$, by the standard asymptotics of Fourier integrals \cite{Wong2001,Economou2006}. At a critical band edge $\omega_{\rm edge}=0$, this corresponds to a branch cut terminating at the origin rather than to an isolated pole.

\subsection{Fractional Schwinger--Dyson expression}
We model the critical correlator by a fractional equation of motion,
\begin{equation}
{}^{C}_{0}D^{\alpha}_{t}\,G(t)+\int_{0}^{t}\Sigma_{\rm eff}(t-t')\,G(t')\,dt'=F(t),
\qquad 0<\alpha<1,
\label{eq:fracEOM}
\end{equation}
where ${}^{C}_{0}D^{\alpha}_{t}$ is the Caputo fractional derivative of order $0<\alpha<1$,
${}^{C}_{0}D^{\alpha}_{t}G(t)=\frac{1}{\Gamma(1-\alpha)}\int_{0}^{t}(t-t')^{-\alpha}\,\dot G(t')\,dt'$~\cite{Caputo1967,Podlubny1999}. Laplace transformation yields
\begin{equation}
\tilde G(s)=\frac{\tilde F(s)+s^{\alpha-1}G(0)}{\,s^{\alpha}+\tilde\Sigma_{\rm eff}(s)\,}.
\label{eq:resolvent}
\end{equation}
Here $F(t)$ is the inhomogeneous source term of the equation of motion. For the
equilibrium correlators studied here the equal-time source is absorbed into the
initial-condition term $s^{\alpha-1}G(0)$ (a property of the Caputo form), and we take
$F(t)=0$ for $t>0$. Note that this should not be confused with the operator equation of motion,
whose source depends on the equal-time (anti)commutator.

The inverse propagator splits into a non-analytic branch-cut term $s^{\alpha}$, which
alone produces $G\sim t^{-\alpha}$ (for constant $\tilde\Sigma_{\rm eff}(s)=\lambda$, i.e.\ a memoryless kernel $\Sigma_{\rm eff}(t)=\lambda\,\delta(t)$, this is the
Mittag--Leffler function $G(0)E_{\alpha}(-\lambda t^{\alpha})\sim t^{-\alpha}$~\cite{SokolovKlafter2002,MetzlerKlafter2000}), and a
residual meromorphic self-energy $\tilde\Sigma_{\rm eff}(s)$ carrying the non-universal, short-time
structure (Fig.~\ref{fig:schematic}, right). The deformation $s\to s^{\alpha}$, the Mittag--Leffler tail, and the
equivalence to a Nakajima--Zwanzig memory kernel $\tilde\kappa_{\rm NZ}(s)\sim
s^{1-\alpha}\mathcal L$, with $\mathcal L$ the generator of the reduced dynamics, were
established for \emph{open} quantum
systems~\cite{PengZhang2025,Ivander2024}, and the singular-plus-regular decomposition of a memory
kernel is the recursion-method / memory-function formalism~\cite{Mori1965,Zwanzig1961,VM1994,Parker2019}.
Here we realize this mechanism as a closed-system, single-particle spectral problem
and, crucially, as a data-driven inverse problem (Sec.~\ref{sec:blind}). 

Two remarks are necessary here; equation~\eqref{eq:fracEOM} is written for $0<\alpha<1$, where the
single initial value $G(0)$ suffices; for $1<\alpha<2$ an additional initial condition
$\dot G(0)$ would enter the Laplace transform. In the operational procedure introduced
below, $\alpha$ is instead used as an \emph{empirical envelope exponent}: the search
grid deliberately includes $\alpha=0$, which corresponds to ordinary pole
extrapolation, and values $\alpha\ge1$, which lie outside the strict fractional form
($\alpha=1$ being the marginal integer case).

\subsection{Why finite-pole fits are ill-conditioned for the exponent on short data}
\label{sec:inseparability}
Near the branch point
$\tilde G(s)\sim s^{\alpha-1}$, and $s^{\alpha-1}$ admits the Stieltjes representation as a
power-law-weighted continuum of poles along the negative real axis,
\begin{equation}
s^{\alpha-1}=\frac{\sin\pi\alpha}{\pi}\int_{0}^{\infty}\frac{\omega^{\alpha-1}}{s+\omega}\,d\omega
\qquad(0<\alpha<1),
\label{eq:stieltjes}
\end{equation}
with spectral weight $\rho(\omega)=\tfrac{\sin\pi\alpha}{\pi}\,\omega^{\alpha-1}$ that diverges as
$\omega\to0$, so a finite-pole estimator with $M$ poles is an $M$-point quadrature of this
continuum.
Over a finite window such a quadrature can be accurate --- a handful of exponentials
reproduces a \emph{gapped, exponentially damped} continuum well~\cite{Erpenbeck2026} --- but
the asymptotic exponent is controlled by the $\omega\to0$ limit of the spectral weight,
which corresponds to $t\to\infty$. Resolving that limit requires poles arbitrarily close to
$\omega=0$; on a finite window only $\mathcal{O}(M)$ nodes are available. Consequently, 
on a finite time window, the branch-point order $\alpha$ of a gapless $\omega=0$
power-law continuum is poorly approximated by a finite-number-pole fit of the raw signal.
The natural remedy is to factor out the cut. We note that the rigorous Laplace-domain
operation, multiplication by $s^{-\alpha}$, corresponds to a fractional
integral (a convolution) in the time domain, \emph{not} to pointwise multiplication by
$t^{\alpha}$. In the real-data procedure below we therefore use a simpler asymptotic
alternative: because the leading envelope scales as $C(t)\sim t^{-\alpha}$, we form
$R(t)=t^{\alpha}C(t)$ and fit the residual with a finite-pole model. This removes the
leading algebraic envelope asymptotically --- precisely the component the finite-pole
model cannot supply --- leaving a residual describable by $\mathcal{O}(1)$ poles, but it
should not be read as exact division by $s^{\alpha}$. This pole-count inflation is visible
in practice: in the DMD forecasting study of Ref.~\cite{Kaneko2025}, a noncritical quench
is captured by $R=13$ modes, whereas an oscillatory correlator with a critical power-law
transient requires $R=193$ modes retained on a window of $10^{4}$ points --- the
finite-pole quadrature tiling the continuum. 

\subsection{ESPRIT as the residual-kernel estimator; rigorous vs. alternative implementations}
\label{sec:esprit}
Since $\tilde\Sigma_{\rm eff}$ is meromorphic, its time-domain image is a sum of
decaying exponentials, $\Sigma_{\rm eff}(t)=\sum_k a_k e^{-\lambda_k t}$ with decay
rates $\mathrm{Re}\,\lambda_k>0$ (Laplace poles at $s=-\lambda_k$, in the stable left
half-plane of Fig.~\ref{fig:schematic}) --- exactly the model that ESPRIT (and Prony /
matrix-pencil) estimates~\cite{Roy1989,Hua1990}. Thus fSD is ESPRIT
augmented by one non-analytic mode $s^{\alpha}$: the fractional order supplies the single
branch-cut basis function that finite exponential sum may hard to represent, and ESPRIT
estimates the meromorphic remainder (Fig.~\ref{fig:schematic}). fSD-ESPRIT therefore extends ESPRIT to the critical (branch-cut) regime that a finite
exponential sum alone is somewhat difficult to reach.

Two implementations are possible. The rigorous route extracts $\Sigma_{\rm eff}$ from the
data through the Dyson relation $\tilde\Sigma_{\rm eff}(s)=s^{\alpha-1}G(0)/\tilde G(s)-s^{\alpha}$
[Eq.~\eqref{eq:resolvent} inverted with $\tilde F=0$],
fits it by a rational model, and forward-integrates the fractional equation. On synthetic
data satisfying Eq.~\eqref{eq:resolvent} we recover the residual poles to $3$--$4$ digits
and extrapolate with errors as small as $10^{-5}$ (Appendix~\ref{app:rigorous}). On the real data generated from
critical systems, however, the Dyson-extracted $\tilde\Sigma_{\rm eff}$ is not well
approximated by a low-order rational and carries right-half-plane poles, so the forward
integration diverges: the clean ``single $s^{\alpha}$ plus meromorphic $\Sigma_{\rm eff}$''
decomposition holds \emph{by construction} for synthetic data but only \emph{approximately}
for real critical dynamics. We therefore use throughout a robust alternative: first detrend the real-time data by
$t^{-\alpha}$, fit the residual with a stabilized finite-pole model, extrapolate, and
finally re-impose the $t^{-\alpha}$ envelope. This alternative method is robust against the imperfect decomposition
because it never divides by the data and its extrapolation is bounded by construction; it
imposes the selected $t^{-\alpha}$ envelope by construction and lets the poles handle
the approach. The method is thus best viewed as a data-selected power-law prior plus a
finite-pole residual model, and its accuracy tests the exponent selector and the
residual fit rather than independently validating the fractional decomposition. This
alternative implementation is the algorithm used for all real-data results below; the rigorous implementation
and its breakdown on real data are examined in Appendix~\ref{app:rigorous}.

\subsection{Blind identification of the critical exponent}
\label{sec:blind}
The detrending order $\alpha$ is itself the critical exponent, and it can be selected from
the fit window alone by holding out validation segments. Here $[t_w,t_D]$ denotes the fit
window: $t_w$ is its start, chosen to exclude the non-universal short-time transient and
fixed to $t_w=3.0$ for the XXZ chain (varied only when estimating window systematics,
Sec.~\ref{sec:uq}), and $t_D$ is the last available time of the data. We split
$[t_w,t_D]$ into a training and a validation segment at three walk-forward training
fractions (the first $55\%$, $65\%$, and $75\%$ of the window), scan $\alpha$, and
choose the value whose detrend-then-pole-fit best predicts the held-out validation
segments, as measured by the median validation error over the three splits; below we
refer to this choice as the residual-minimization selector. No closed-form value and no out-of-window data
enter. Note that this procedure differs from both existing routes to an exponent. A finite-pole
estimator such as ESPRIT returns a set of discrete complex poles, and the
algebraic tail $t^{-\alpha}$ is a property of the $\omega\to0$ continuum as a whole, not
the decay rate of any single exponential. In the open-system fractional framework of
Ref.~\cite{PengZhang2025}, on the other hand, $\alpha$ is an input rather than an output:
the fractional order is fixed beforehand by the known spectral exponent of the bath, so
the power law is predicted from the physics without reference to the time-domain data.
In our work, $\alpha$ is treated as the unknown
of an inverse problem and is read off from the short-time data themselves.

\begin{figure*}[t]
\includegraphics[width=0.80\linewidth]{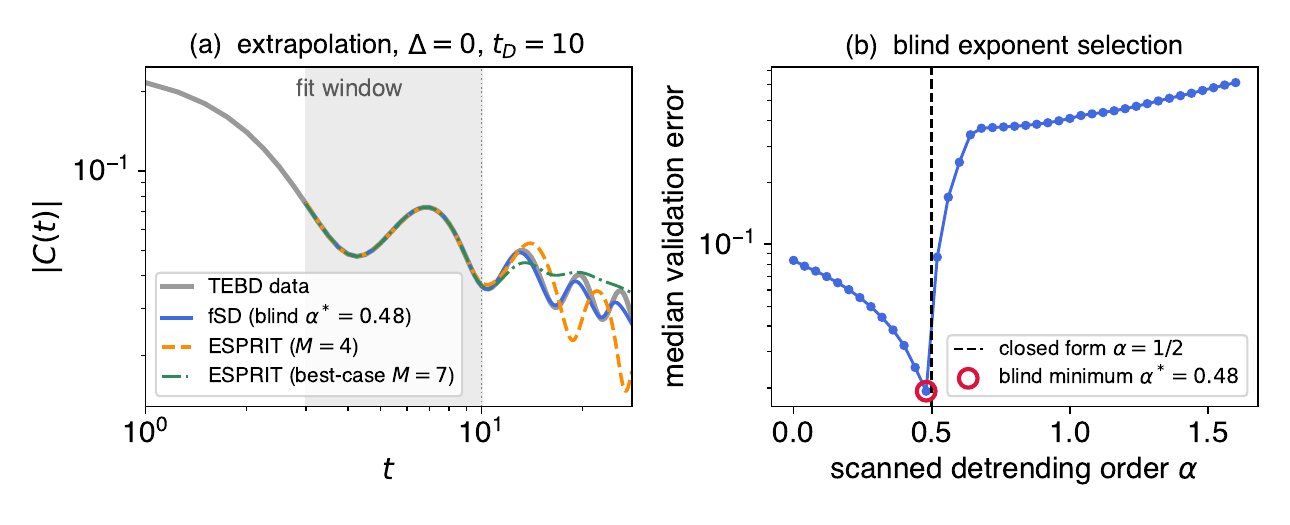}
\caption{\label{fig:method}
An example comparison between the fSD and ESPRIT on the XXZ chain at $\Delta=0$ (and $h=0$). (a) $|C(t)|$ from TEBD
(gray), with the fSD (blue solid, blind $\alpha^{*}$) and finite-pole ESPRIT (orange
dashed) extrapolations fitted on the short window $t\in[3,10]$ (shaded, identical
settings and pole count $M=4$ for both); beyond the window the finite-pole fit departs
from the data, while fSD tracks the algebraic tail (out-of-window error $0.07$ vs.\
$0.54$). The green dash-dotted curve shows ESPRIT at its best-case pole count ($M=7$,
chosen with knowledge of the exact tail, App.~\ref{sec:protocol}): the error improves to
$0.19$ but remains $2.7\times$ that of fSD. (b) Blind exponent selection: median
walk-forward validation error (three training fractions, App.~\ref{sec:protocol})
versus the scanned detrending order $\alpha$. The minimum $\alpha^{*}=0.48$ (circle) lies next to the
closed-form Luttinger value $\alpha=1/2$ (dashed line), which enters neither the fit nor
the selection. All fits use the full complex $C(t)$; quoted errors are evaluated on
$\mathrm{Re}\,C$ [Eq.~(\ref{eq:oow})], of which the displayed $|C(t)|$ is the envelope.}
\end{figure*}

Figure~\ref{fig:method} demonstrates the whole procedure on the XXZ chain at $\Delta=0$ (and $h=0$)
with a short fit window, $t\in[3,10]$. Panel (a) shows the raw correlator together with
the fSD and finite-pole (ESPRIT) extrapolations: inside the fit window the two fits
are equally good, but beyond it the finite-pole extrapolation departs from the data,
while the detrended fit follows the algebraic tail out to $t=28$. Increasing the pole
count does not close the gap: for $M\ge5$ the window itself is fitted essentially
exactly, yet the out-of-window error saturates near $0.2$, and even at the best-case
pole count ($M=7$, selected using knowledge of the exact tail, App.~\ref{sec:protocol})
it remains three times that of fSD. More poles improve the interpolation, not the
extrapolation. Panel (b) shows the
blind selection at work: the median walk-forward validation error develops a sharp minimum at
$\alpha^{*}=0.48$, right next to the closed-form value $\alpha=1/2$ (dashed line), which
enters neither the fit nor the selection.

\section{Methods}
\subsection{Models and tensor-network data}
We study two $T=0$ systems. (i) The spin-$\tfrac12$ XXZ chain
$H=\sum_i[S^x_iS^x_{i+1}+S^y_iS^y_{i+1}+\Delta S^z_iS^z_{i+1}]$ at zero field on the
critical line $-1<\Delta\le1$ ($z=1$ Luttinger liquid), with closed-form decay exponent
$\alpha(\Delta)=1-\arccos(\Delta)/\pi$ --- the leading (staggered) Luttinger-liquid
exponent $1/(2K)$, where $K=\pi/\{2[\pi-\arccos\Delta]\}$ is the Luttinger parameter
of the XXZ chain~\cite{LutherPeschel1975,Haldane1980,Giamarchi2004}, with uniform and higher-harmonic components subleading over the
windows analyzed here, and with a marginal logarithmic correction at
$\Delta=1$~\cite{AGSZ1989,GiamarchiSchulz1989}; the local autocorrelation $C(t)=\langle
S^x_i(t)S^x_i\rangle$ is computed by quantum-number-conserving time-evolving block decimation
(TEBD)~\cite{Vidal2003,Vidal2004}, as implemented in the ITensor
library~\cite{Fishman2022} ($N=100$, $\delta t=0.05$, to $t_{\max}=28$). All chains use
open boundary conditions with the measured site at the chain center; time evolution
uses a second-order Trotter decomposition with truncation cutoffs of
$10^{-10}$--$10^{-11}$ and bond-dimension caps between $512$ and $1024$ depending on
the data set. 
For the
magnetic-field scan of Sec.~\ref{sec:hscan} we add a Zeeman term $-h\sum_i S^z_i$ at
$\Delta=1$ (saturation field $h_{\rm sat}=1+\Delta=2$) and use TEBD data sets with
$N=160$ [$h=2.0$, $2.1$ with $t_{\max}=100$; $h=1.7$, $1.9$ from a
quantum-number-conserving ground-state search ($t_{\max}=32$ and $58$), see
Sec.~\ref{sec:hscan}], $N=100$ ($t_{\max}=200$; $h=3.0$), and $N=50$ ($t_{\max}=30$;
$h=1.0$, $1.5$, $2.5$); out-of-window errors are always evaluated within the
reflection-free time region. (ii) The single-orbital Hubbard model on the Bethe lattice solved by
single-site DMFT with a fork tensor-network impurity solver~\cite{Bauernfeind2017},
whose time evolution is performed with the two-site time-dependent variational principle
(TDVP)~\cite{Haegeman2011,Haegeman2016}. The model is studied at half filling
(particle-hole symmetric), where the noninteracting density of states is semicircular,
$\rho_{0}(\omega)=\tfrac{2}{\pi D^{2}}\sqrt{D^{2}-\omega^{2}}$, with the half-bandwidth
$D$ setting the unit of energy.
The retarded $G^R(t)$ is obtained to $t_{\max}=30$ across $U/D\in\{2.0,\dots,4.0\}$. The DMFT
complex-time$\to$real-time reconstruction follows Refs.~\cite{Cao2024,Grundner2024,Yu2026,ChaLeeKim2025}
and is not modified here. Throughout the analysis the two data sets are treated
identically: the XXZ autocorrelation $C(t)$ and the DMFT retarded Green's function
$G^{R}(t)$ each play the role of the generic correlator $G(t)$ in the fSD equations
of Sec.~\ref{sec:theory}.

\subsection{Baselines and uncertainty quantification}
\label{sec:uq}
We compare against three finite-pole / sum-of-exponential extrapolators on the same
input: a time-domain linear-prediction (LP) extension~\cite{WhiteAffleck2008} in the
spirit of Ref.~\cite{Kemper2024} (we do not reproduce their
positive-semidefinite-Hankel denoising and label it accordingly), ESPRIT / matrix
pencil~\cite{Roy1989,Hua1990}, and an adapted Erpenbeck-style
extrapolator~\cite{Erpenbeck2026} (ESPRIT with
$L=0.4N$, the mode number set by their singular-value threshold $s_i/s_0>10^{-6}$ as in
the noiseless benchmark of Ref.~\cite{Erpenbeck2026}, exponentially growing modes
discarded, amplitudes by Vandermonde least squares; we omit their
smallest-exponent-to-zero step, which targets a nonzero steady state inappropriate for
the decaying critical signal). Rational-approximation methods such as
AAA~\cite{AAA2018} belong to the same finite-pole class and face the same conditioning
limitation (Sec.~\ref{sec:inseparability}). The DMD recently
applied to forecasting quantum many-body dynamics~\cite{Kaneko2025} belongs to the same
estimator family (it is essentially a matrix-pencil method), so the ESPRIT baseline
represents it here. For all fSD extrapolations we follow the procedure described in Sec.~\ref{sec:esprit}. 
All quantitative comparisons between fSD and the finite-pole baselines are run on the
same fit window $[t_w,t_D]$ and the same subsampled time grid. For a fit cutoff $t_D$ the out-of-window error is computed in the interval $t_D < t_j < t_{\rm out}$ as follows,
\begin{equation}
\label{eq:oow}
E(t_D)=\frac{\big[\sum_{t_j}\,[\mathrm{Re}\,C_{\rm pred}(t_j)-\mathrm{Re}\,C_{\rm ref}(t_j)]^{2}\big]^{1/2}}
{\big[\sum_{t_j}\,[\mathrm{Re}\,C_{\rm ref}(t_j)]^{2}\big]^{1/2}}.
\end{equation}
For more computational details see App.~\ref{sec:protocol}.

Two blind estimators of the exponent appear in this work: the residual-minimization selector of Sec.~\ref{sec:blind}, which supplies the $\alpha$ actually used in the extrapolation pipeline (for Fig.~\ref{fig:pillarB} it is run with the identical pipeline settings --- subsampled grid, automatic $M$ --- at the central window $(t_w,t_D)=(3.0,20)$, with the window systematic taken over $t_w\in\{2.5,3.0,3.5\}$ and $t_D\in\{16,18,20,22\}$), and the envelope estimator described here, which provides an independent measurement of the same quantity. The envelope estimator fits a power
law to the rolling-maximum envelope of $|C(t)|$ (the running maximum over a sliding
window of 31 samples, evaluated every 16 samples) by linear regression of the
logarithm of the envelope versus $\ln t$ over $t\in[2.5,\,t_{\rm hi}]$; here
$t_{\rm hi}$ is the upper end of the envelope fit range, set to $t_{\rm hi}=24$ unless
stated otherwise.

Uncertainty on the blind exponent is quantified with two error sources, statistical and
systematic. The statistical error applies naturally to the envelope estimator, which is
a linear regression: its error bar combines in quadrature the regression standard error
with a jackknife estimate, obtained by deleting one envelope sample at a time, refitting
the power law to the remaining samples, and converting the spread of the refitted
exponents into a standard error. The $\alpha$-grid-search selector
(Sec.~\ref{sec:blind}), by contrast, carries no natural statistical error, being the
minimizer of a discrete scan. The systematic error is the window dependence and applies
to both estimators: the entire analysis is repeated over a set of window choices
$(t_w,t_D,t_{\rm hi})$, and the spread of the resulting exponents is quoted as the
window-systematic error bar --- for the grid-search selector this is the only error bar
quoted. Because the envelope samples come from overlapping rolling windows and the
underlying TEBD data are deterministic, the regression-plus-jackknife bar should be
read as an internal stability measure rather than an independent statistical error;
the dominant uncertainty is the window dependence.

\section{Results: XXZ critical line}
\subsection{Data-efficient extrapolation of the critical tail}
\label{sec:dataeff}

\begin{figure*}[t]
\includegraphics[width=0.99\linewidth]{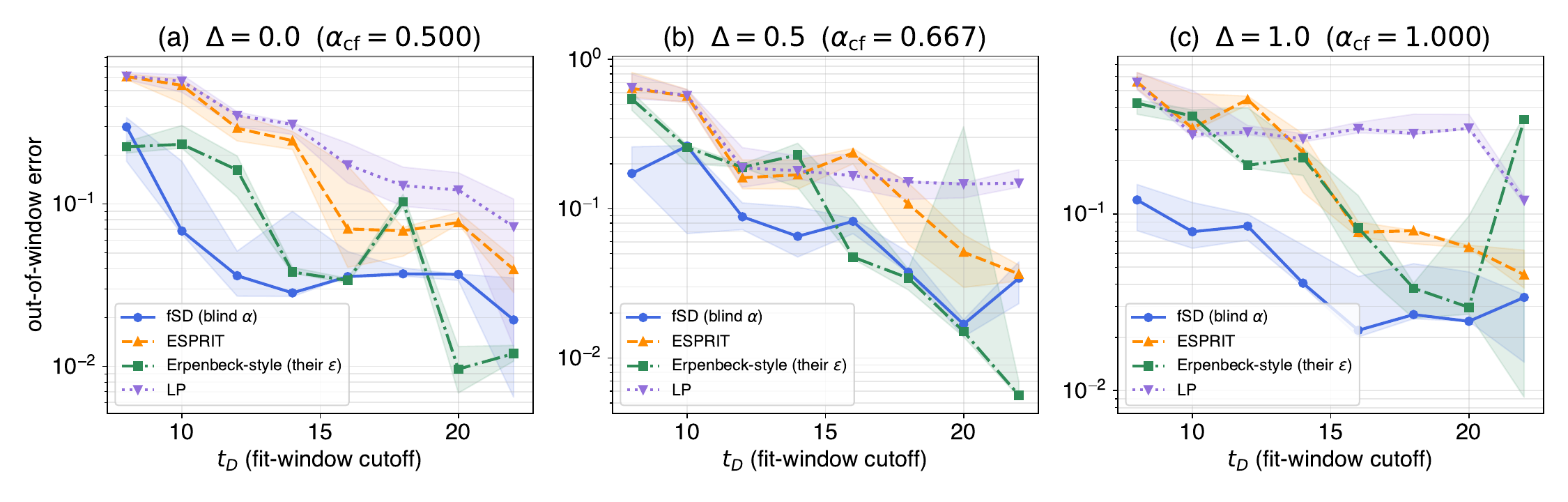}
\caption{\label{fig:dataeff}
Data efficiency under a fully symmetric, blind procedure. Out-of-window extrapolation
error versus fit-window cutoff $t_D$ for the XXZ critical line at $\Delta=0$ (a),
$0.5$ (b), and $1$ (c), comparing fSD (blue; residual fit with auto-selected pole count, $\alpha$ identified
blind from the fit window alone) with three finite-pole baselines on the same data and
the same grid: plain ESPRIT (orange, shared auto-$M$), the Erpenbeck-style extrapolator
of Ref.~\cite{Erpenbeck2026} run with its authors' prescriptions ($L=0.4N$, mode number
from their singular-value threshold $10^{-6}$, growing modes discarded, Vandermonde
amplitudes; green), and a time-domain LP extension in the spirit of
Ref.~\cite{Kemper2024} (purple, shared auto-$M$). Lines are medians over the fit-window
start $t_w\in\{2.5,3.0,3.5\}$; shaded bands show the corresponding $t_w$-systematic
spread. In the short-data regime the branch-cut ansatz is the most accurate in ten of
twelve cases (median $2.9\times$ over the full remaining window; $2.5\times$ at fixed
prediction horizon $H=8$ and $3.1\times$ at $H=16$, App.~\ref{sec:protocol}).}
\end{figure*}

We first examine how accurately the critical tail can be extrapolated from a short data
window. Figure~\ref{fig:dataeff} presents the out-of-window extrapolation error as a
function of the fit-window cutoff $t_D$ for the XXZ chain at $\Delta=0$, $0.5$, and $1$,
comparing fSD with the three finite-pole baselines of Sec.~\ref{sec:uq}: plain ESPRIT,
the Erpenbeck-style ESPRIT extrapolator~\cite{Erpenbeck2026}, and the time-domain LP
extension in the spirit of Ref.~\cite{Kemper2024}. For each $t_D$, all four methods are
fitted on the same window $[t_w,t_D]$ under the procedure of
App.~\ref{sec:protocol}, and the extrapolations are
compared against the reference data beyond $t_D$. Solid lines are medians over the fit-window starting points
$t_w\in\{2.5,3.0,3.5\}$, and the shaded bands show the corresponding spread --- the same
$t_w$-systematic treatment for every method alike.

Two features in Fig.~\ref{fig:dataeff} are worth noting. First, the advantage tends to be largest where
in the short data window regime --- relative to plain ESPRIT it reaches almost an order of
magnitude for $t_D\le12$ (e.g.\ $7.9\times$ at $\Delta=0$, $t_D=10$) --- which is
precisely the regime relevant in practice, where entanglement growth limits the
accessible simulation time. Second, the advantage decays as the window grows: the ratios
fluctuate from window to window, and from $t_D\approx16$ the Erpenbeck-style baseline
overtakes at $\Delta=0$ and $0.5$, while at $\Delta=1$ a residual advantage persists to
the longest window studied. We
emphasize that this crossover is part of the result rather than a shortcoming: once the
window is long enough for the finite-pole quadrature of Sec.~\ref{sec:inseparability} to
converge, nothing further is gained by treating the branch cut separately. 

The Erpenbeck-style baseline deserves a separate comment, because it is not run with
the shared auto-$M$ rule but with the original prescriptions of
Ref.~\cite{Erpenbeck2026} (see Sec.~\ref{sec:uq}).
%
%
Because the data are
essentially noiseless, the small threshold retains substantially more modes than the
noise-conservative auto-$M$ rule, and the baseline behaves accordingly. It improves
markedly on plain ESPRIT at many windows, but at the price of a much larger
window-to-window variance: it is the best method on a few short windows and
occasionally the worst, as at $\Delta=1$, $t_D=22$. The qualitative picture is
nevertheless unchanged. Over the twelve short-window cases ($t_D\le14$), fSD remains
the most accurate method in ten, with a median advantage of approximately $2.9\times$
over the best finite-pole variant at each point, and it remains the best or comparable
with the best across the entire range at $\Delta=1$. The fixed-horizon metric of
App.~\ref{sec:protocol} confirms that this advantage is not an artifact of the
shrinking evaluation window: over the same twelve cases the median advantage is
$2.5\times$ at $H=8$ (eleven of twelve cases, where the error is estimated in the interval of [$t_D$, $t_D + H$]; see Sec.~\ref{sec:uq}) and $3.1\times$ at $H=16$, narrowing to
$1.2\times$ only at the shortest horizon $H=4$ --- the benefit of the algebraic
envelope grows with the prediction distance, as expected for an asymptotic form. On long windows the adapted
baseline overtakes fSD from $t_D\approx16$ at both $\Delta=0$ and $\Delta=0.5$, while
against plain ESPRIT the advantage persists across most of the range studied (parity is
reached at $\Delta=0.5$ only by $t_D\approx22$). This is consistent with the
quadrature picture of Sec.~\ref{sec:inseparability}: given enough data, more retained
poles tile the continuum better.

\subsection{Blind identification of $\alpha(\Delta)$ with quantitative uncertainty}
\label{sec:pillarB}

\begin{figure}[t]
\includegraphics[width=1.0\linewidth]{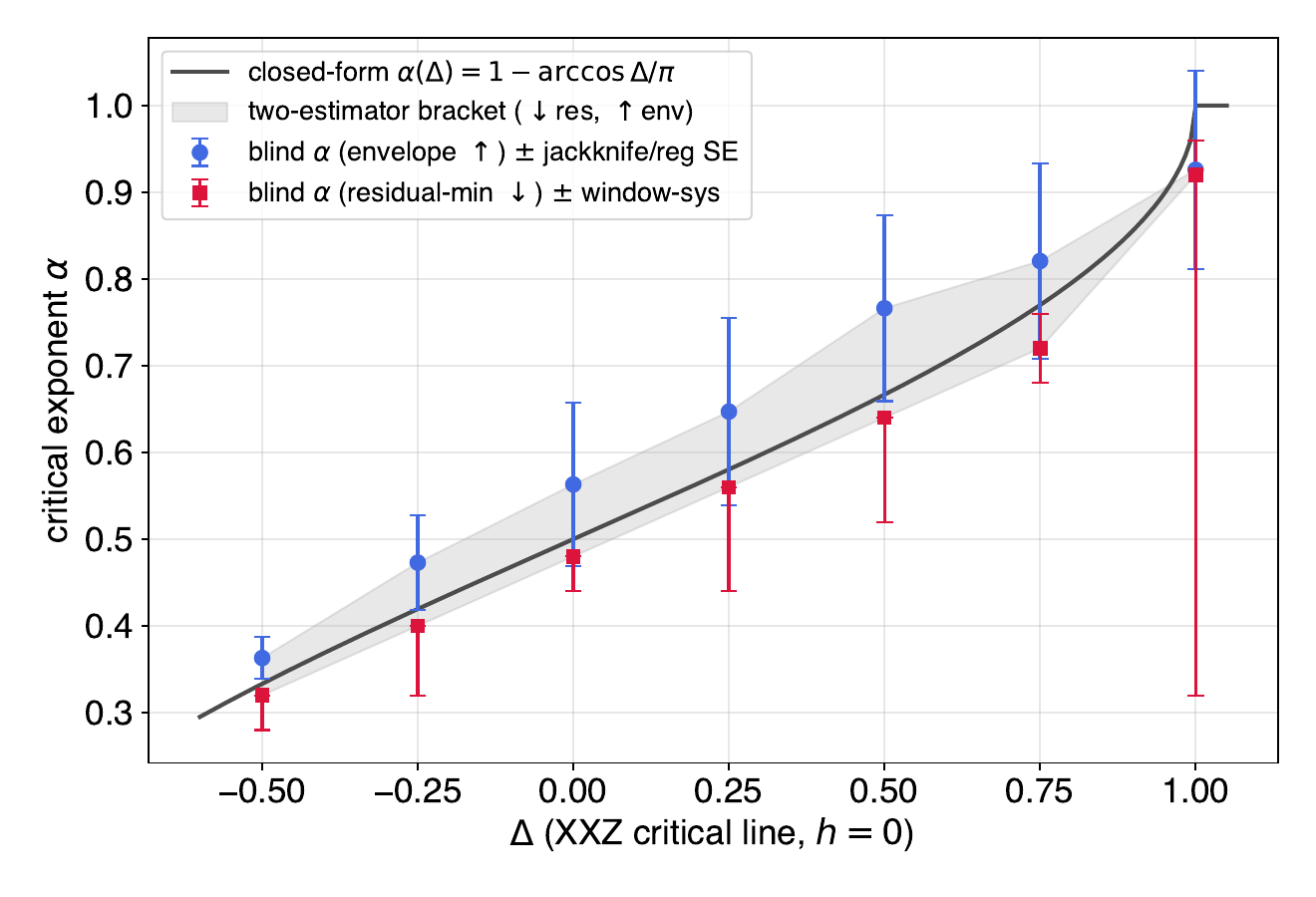}
\caption{\label{fig:pillarB}
Blind identification of the critical exponent. Blind estimates of $\alpha$ along the
XXZ critical line compared with the closed-form Luttinger value (gray, not used in the
identification). Two independent blind estimators --- envelope power-law fit (blue, with
jackknife/regression error bars) and residual-minimization (red, with
window-systematic bars) --- bracket the closed form across $\Delta\in[-0.5,0.75]$ (shaded).
At $\Delta=1$ both estimators undershoot, reflecting the known bias from the
multiplicative logarithmic correction $t^{-1}(\ln t)^{1/2}$ (see text).}
\end{figure}

We next turn to the identification of the critical exponent itself. In the fSD
pipeline the detrending order $\alpha$ is not an auxiliary fitting parameter: it \emph{is}
the critical exponent of the spectral function, $A(\omega)\sim|\omega|^{\alpha-1}$, and it
is selected from the fit window alone. Figure~\ref{fig:pillarB} compares the blindly
identified exponent with the closed-form Luttinger value
$\alpha(\Delta)=1-\arccos(\Delta)/\pi$~\cite{LutherPeschel1975,Haldane1980} along the critical line, using two independent
blind estimators: the residual-minimization selector of Sec.~\ref{sec:blind} and the
envelope power-law fit of Sec.~\ref{sec:uq}. Across $\Delta\in[-0.5,0.75]$ the two
estimators bracket the closed-form value, with the envelope estimator carrying a
quantitative error bar of $\pm0.02$--$0.11$ (statistical and window-systematic contributions combined in quadrature). The
residual-minimization selector is additionally limited by its finite search grid to a
quantization resolution of half the grid spacing ($\pm0.02$), which is comparable to
its smallest window-systematic bars and sets a floor on the quoted precision. 
At the Heisenberg point $\Delta=1$ both estimators yield lower values than the
exact one. This is not a statistical fluctuation but a bias of known origin: at
$\Delta=1$ the correlator decays as $t^{-1}(\ln t)^{1/2}$~\cite{AGSZ1989,GiamarchiSchulz1989}, and fitting a pure power law
to a signal carrying a multiplicative logarithmic correction systematically
underestimates the exponent. The practical significance of the blind identification is
that it makes the extrapolation of Sec.~\ref{sec:dataeff} fully self-contained: the
exponent used for detrending is obtained from the same short-time data that are being
extrapolated, with no external model input.

\subsection{Magnetic-field scan at $\Delta=1$: a $z=2$ critical point and a gapped phase}
\label{sec:hscan}

\begin{figure*}[t]
\includegraphics[width=0.99\linewidth]{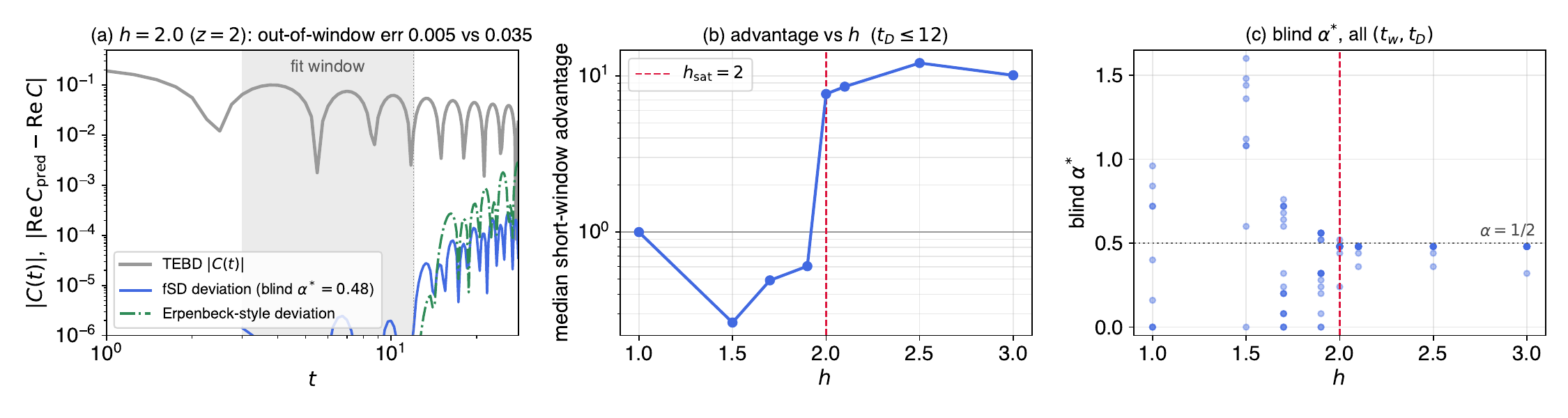}
\caption{\label{fig:hscan}
Magnetic-field scan at $\Delta=1$ under the same fully blind procedure as
Fig.~\ref{fig:dataeff}. (a) Real-time comparison at the saturation field $h=2.0$:
$|C(t)|$ from TEBD (gray) and the deviations $|\mathrm{Re}\,C_{\rm pred}-\mathrm{Re}\,C|$
--- evaluated on the real part, matching the error metric of Eq.~(\ref{eq:oow}) --- of fSD (blue, blind
$\alpha^{*}=0.48$) and of the Erpenbeck-style baseline (green), both fitted on
$t\in[3,12]$ (shaded). The small in-window fSD residual reflects its conservative
auto-$M$ pole budget ($M=4$ vs.\ $6$ retained by the baseline's $10^{-6}$ threshold);
the baseline interpolates the window three orders of magnitude more closely yet
extrapolates an order of magnitude worse --- the interpolation-versus-extrapolation
contrast of Fig.~\ref{fig:method}. (b) Median short-window ($t_D\le12$) advantage of fSD over the
best of the three finite-pole baselines versus $h$; the dashed line marks
$h_{\rm sat}=2$. (c) Blind exponent $\alpha^{*}$ for every $(t_w,t_D)$ pair versus $h$: for
$h\ge2$ --- at the $z=2$ saturation point and in the gapped phase --- $\alpha^{*}$
locks onto the dilute-magnon value $1/2$ (dotted line) for every window with
$t_D\ge12$, while below saturation
[$h\le1.5$, and $h=1.7$, $1.9$ with quantum-number-conserving ground states] it
scatters or drifts (see text).}
\end{figure*}

So far the critical line has been scanned in the anisotropy $\Delta$ at zero field,
where the low-energy theory is always a $z=1$ Luttinger liquid. A magnetic field at
$\Delta=1$ provides a stringent and more varied test, because it drives the chain
through three qualitatively different regimes: the finite-magnetization Luttinger
liquid for $h<h_{\rm sat}$; the dilute-magnon saturation transition at
$h_{\rm sat}=2$ --- a critical point in a universality class different from the
conformal $z=1$ class of the preceding sections, with dynamical exponent $z=2$ and
spectral edge $A(\omega)\sim\omega^{-1/2}$, i.e.\ $\alpha=1/2$~\cite{Sachdev2011}; and the fully polarized
gapped phase for $h>h_{\rm sat}$, whose single-magnon excitations --- precisely the
states probed by $C(t)$ --- are exactly free. We repeat the analysis of
Secs.~\ref{sec:dataeff} and~\ref{sec:pillarB} on this scan with the identical blind
procedure.

Figure~\ref{fig:hscan} summarizes the outcome. At the saturation field and throughout
the gapped phase ($h=2.0$--$3.0$), the blind exponent locks onto the dilute-magnon
value, $\alpha^{*}=0.48$ for every window with $t_D\ge12$ [panel~(c)], and fSD
extrapolates the short-window data $7.7$--$12\times$ more accurately (median over
$t_D\le12$) than the best of the three finite-pole baselines [panel~(b)]. At the
saturation field itself this is precisely the $z=2$ prediction: with no model input,
the blind selection identifies $\alpha=1/2$, demonstrating that the identification is
not tied to the conformal $z=1$ class of the zero-field scan. Below saturation
($h=1.7$, $1.9$), by contrast, the short-window exponent does not lock but scatters and
drifts, and the short-window advantage disappears ($0.5$--$0.6\times$) --- the
finite-magnetization Luttinger liquid is multi-component (see below), and the boundary
between the two behaviors sits exactly at $h_{\rm sat}$. Panel~(a) shows the real-time comparison
at $h=2.0$ in the format of Fig.~\ref{fig:method}(a): beyond the fit window the
deviation of the Erpenbeck-style baseline grows to the percent level, while the fSD
deviation stays an order of magnitude smaller (out-of-window errors $0.005$ vs.\
$0.035$).

Two further observations are instructive --- one positive, one negative. First, the
advantage \emph{persists in the gapped phase} ($h=2.1$--$3.0$), where one might naively
expect a pole description to suffice. The reason is that the magnons of the polarized
phase are exactly free: the band-edge van~Hove singularity $A(\omega)\propto(\omega-\omega_{\rm edge})^{-1/2}$ is perfectly sharp, so the
correlator follows asymptotically $C(t)\propto e^{-i\omega_{\rm edge}t}t^{-1/2}$, with an undamped algebraic envelope $\sim t^{-1/2}$ multiplying the
band-edge oscillations \cite{Wong2001,Economou2006} (which poles in a finite-time window are hard to represent) and the
blind selection duly finds $\alpha^{*}\approx1/2$ here as well. 
In fact, for $h\ge h_{\rm sat}$ the correlator is known in closed form: $S^{x}_{i}$ acting on the polarized ground state creates a single magnon, whose dispersion $\epsilon(k)=h-\Delta+\cos k$ is exact, so that $C(t)=\frac{1}{4}\,e^{-i(h-\Delta)t}J_{0}(t)$ in the thermodynamic limit~\cite{Economou2006}, with the asymptotics $J_{0}(t)\simeq\sqrt{2/\pi t}\,\cos(t-\pi/4)$ making the undamped $t^{-1/2}$ envelope and the band-edge oscillation explicit; the TEBD data reproduce this closed form throughout the reflection-free window (not shown), so that the gapped-phase and saturation-point results are anchored to an exact benchmark. As discussed in
Sec.~\ref{sec:falsify}, this sharpens the scope of the method: the operative criterion is
the presence of sharp algebraic spectral singularities, rather than gaplessness itself.

Second, for $h\le1.5$ --- the finite-magnetization Luttinger liquid --- the blind
exponent does not stabilize and no significant advantage survives. The dominant reason
is physical rather than statistical. At finite magnetization the local transverse
correlator at finite time is a superposition of several critical components with different exponents
and subleading harmonics~\cite{Giamarchi2004,HikiharaFurusaki2004}, and a single detrending
order cannot represent two comparable algebraic envelopes at once. The walk-forward
validation loss makes this quantitative: at $h=2.0$ it drops by more than two orders of
magnitude into a sharp minimum at $\alpha=0.48$ (near-optimal width
$\Delta\alpha\approx0.04$), whereas at $h=1.0$ the landscape is shallow and broad
(near-optimal width $\Delta\alpha\gtrsim1$, several local minima) and its best value
remains $0.39$ --- even the optimal single-order detrend leaves unmodeled structure, so
the selected $\alpha^{*}$ wanders with the window. Importantly, the failure is
self-diagnosing rather than silent (exactly as for the pole-dominated DMFT data analyzed in
Sec.~\ref{sec:falsify}); the scatter of $\alpha^{*}$ across windows itself signals that
a single algebraic envelope is not the right model, and since $\alpha=0$ belongs to the
search grid, the blind selection can always return the plain pole fit.
A complementary field scan at
$\Delta=0.5$, on data generated entirely through the complex-time route, shows the same
physics from the low-field side and is presented in Appendix~\ref{app:d05}.

\section{Results: falsifiability test on single-orbital DMFT}
\label{sec:falsify}

\begin{figure*}[t]
\includegraphics[width=0.80\linewidth]{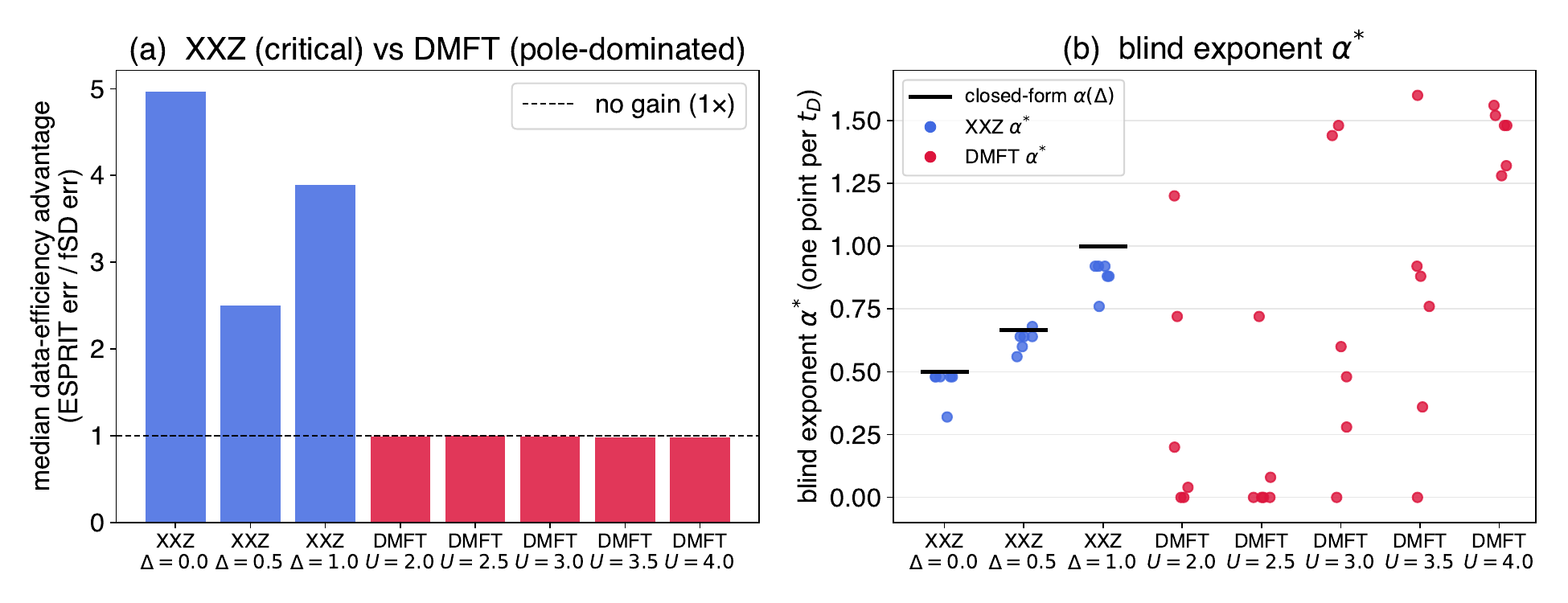}
\caption{\label{fig:falsify}
Falsifiability on a pole-dominated system. (a) Data-efficiency advantage of fSD,
defined per system as the median over the fit windows $t_D\in\{8,10,12,14,18,22\}$
($t_w=3.0$) of the ratio of the plain-ESPRIT error to the fSD error (evaluated on the
complex modulus for the purely imaginary DMFT signal; App.~\ref{sec:protocol}): $2.5$--$5\times$ on
the gapless XXZ critical line, $\sim\!1\times$ on the pole-dominated DMFT
($U/D=2$--$4$). Plain ESPRIT (shared auto-$M$ and stabilization) is chosen for the comparison here
because the ratio then isolates the effect of the fractional degree of freedom; the
Erpenbeck-style baseline, which differs also in its mode budget, is compared in
Fig.~\ref{fig:dataeff}. The fSD exponent $\alpha$ is selected
blind for every window, on both data sets (XXZ and DMFT). (b) The blind exponent $\alpha^{*}$ (one
point per $t_D$) clusters near the closed-form $\alpha(\Delta)$ (black bars) for XXZ but
scatters over the search range $[0,1.6]$ with no stable value for DMFT (see text).}
\end{figure*}

We now turn to the single-orbital Hubbard model on the Bethe lattice --- a noncritical
control whose low-frequency response is regular. The real-time impurity Green's
function $G^R(t)$ is reconstructed by extrapolating complex-time contours to the real
axis, as established in Refs.~\cite{Cao2024,Yu2026} and benchmarked for spin chains in
our earlier work~\cite{ChaLeeKim2025}; the reconstruction agrees with the direct
real-time evolution to within $\sim3\times10^{-4}$ and reaches $t=30$ in the metallic,
transitional, and insulating regimes. The spectral functions obtained from these data
--- and the resolution gained by extrapolating them to long times --- are collected in
Appendix~\ref{app:dmftAw}; here we use the time-domain signals directly.

A method built around a power-law ansatz must also be tested against the opposite failure
mode: fitting a power law where none exists. The single-orbital Hubbard model on the Bethe lattice provides exactly this test,
because its spectrum consists of a quasiparticle peak and Hubbard bands --- poles, in the
language of Sec.~\ref{sec:theory} --- with no gapless power-law continuum.
Figure~\ref{fig:falsify} summarizes what happens when the identical blind pipeline is
applied to both systems; in particular, the exponent is \emph{not} fixed to zero on the
DMFT data --- $\alpha$ is selected blind for every window, exactly as on the XXZ data.
Panel (a) shows the median data-efficiency advantage: $2.5$--$5\times$ on the gapless XXZ
critical line, but only $\sim\!1\times$ (no advantage) on the DMFT data
across $U/D=2$--$4$. Panel (b) provides a sharper diagnostic:
on the XXZ line the blind exponent $\alpha^{*}$ clusters near the closed-form
$\alpha(\Delta)$ for every data length, whereas on the DMFT data it scatters over the
entire search range $[0,1.6]$ without settling on any stable value: when the data contain
no branch cut, the blind selection finds no preferred fractional order, and fSD
reduces to the standard pole result instead of manufacturing a spurious power law. This
behavior is what makes the advantage on the critical line trustworthy.

Taken together, the XXZ and DMFT results show when fSD performs well,
and the field scan of Sec.~\ref{sec:hscan} makes the criterion precise. The operative
condition is the presence of a sharp algebraic singularity in the spectral
function --- a power-law branch point that is not smoothed by a finite lifetime --- so
that the correlator carries an algebraic envelope persisting indefinitely in time. A
gapless $\omega=0$ continuum is the canonical example (the zero-field critical line and
the $z=2$ saturation point), but the sharp band-edge van~Hove singularities of the
free-magnon phase act in exactly the same way even though the spectrum there is gapped
(Sec.~\ref{sec:hscan}): in both cases the associated Stieltjes pole continuum cannot be
tiled by finitely many poles on a short window (Sec.~\ref{sec:inseparability}).
Conversely, when interactions endow the excitations with finite lifetimes, the
singularities are broadened: the spectral function becomes, to the accessible accuracy,
a finite collection of resonances, its time-domain image a finite sum of damped
exponentials, and a finite-pole fit is then the right model. This is the situation in
the DMFT solution of the Hubbard model --- the quasiparticle peak and the Hubbard bands
are lifetime-broadened by the self-energy --- and it is why the blind exponent finds no
stable value there and fSD reduces to the standard pole result
(Fig.~\ref{fig:falsify}). In addition, the comparison of the two gapped cases makes the point
clearer: a Mott-insulator band edge and a polarized-phase magnon band edge have a similar
analytic structure in the complex plane, yet fSD gains nothing on the former (damped)
while it outperforms the pole fits severalfold on the latter (sharp).


\section{Discussion and Summary}
In the Laplace plane, a gapless,
undamped power-law continuum corresponds to a continuum of poles whose density diverges
at the origin; on a finite data window, the branch-point exponent is therefore
ill-conditioned for a finite-pole fit. The data-selected envelope supplies precisely
this poorly representable component and delegates everything else to a standard pole
fit. Consistent with this
picture, the benefit appears precisely where entanglement growth forces a short window,
and it disappears once the window becomes long (the crossover in Fig.~\ref{fig:dataeff}). 

Our approach is complementary to existing techniques rather than a replacement for them.
Complex-time analytic continuation~\cite{Cao2024,Grundner2024,Yu2026,ChaLeeKim2025}
operates on the same type of input and is highly effective in the pole-dominated regime;
notably, those works report that an ad-hoc broadening is required at Hubbard band edges
``difficult for the complex-pole representation to describe'' --- direct evidence for
the poles-versus-branch-cut distinction drawn here. 
The
$s^{\alpha}$ fractional mechanism and the Mittag--Leffler tail were developed for open
quantum systems~\cite{PengZhang2025}, and the finite-pole long-time
extrapolators~\cite{Erpenbeck2026,Kemper2024} are the appropriate tools when the spectrum
is a finite collection of poles, but they do not address a non-analytic tail. The
successful DMD forecasts of Ref.~\cite{Kaneko2025} operate in exactly this regime: a
long, dense input window ($10^{4}$ points over $t\le100$) with up to $R\approx200$
retained modes, and an infinite-time limit that is itself a constant (a pole at
$\omega=0$); this is the data-rich side of the crossover identified here. The contribution of the present work is to combine these
elements into a closed-system, data-driven inverse problem: the branch-cut order is
identified from the short-time data and then used to extrapolate the same data.

Several limitations should be stated explicitly. (i) The demonstrated advantage is
confined to the short-data regime: for a system that can be simulated to long times, a
pole method applied to the full window is sufficient. The practical value of the data
efficiency therefore lies in harder critical problems where $t_{\max}$ is intrinsically short. (ii) The clean decomposition
into $s^{\alpha}$ plus a low-order meromorphic residual holds by construction only for
synthetic data; on real data the rigorous forward integration is unstable
(Appendix~\ref{app:rigorous}), and we rely on the detrend-and-refit proxy, which is
justified operationally rather than derived; a stable, passive implementation of the
fully Dyson-extracted closure remains an open algorithmic task. (iii) At $\Delta=1$ both blind estimators
carry the bias from the multiplicative logarithmic correction $t^{-1}(\ln t)^{1/2}$
discussed above. (iv) At finite magnetization below saturation ($\Delta=1$, $h\le1.5$) several
critical components with different exponents coexist and the single-order blind
identification does not stabilize (Sec.~\ref{sec:hscan}); a two-order extension of the
closure is the natural remedy. 

A promising extension is to strongly correlated metallic regimes with scale-invariant local dynamics. Within DMFT, power-law single-particle spectra have been found near the Mott-critical regime of the one-band Hubbard model, including an asymptotic scale-invariant solution and a broader crossover regime associated with approximate local quantum criticality \cite{Eisenlohr2019}. More recently, asymptotic power-law spectra and $\omega/T$ scaling in single-particle, spin, and charge responses were reported in the orbital-selective Mott phase of a three-orbital Hubbard model \cite{Eickhoff2026}. The corresponding real-time correlators are expected to exhibit long-lived algebraic envelopes, making them natural targets for the blind fractional-envelope reconstruction developed here. Applying the method directly within real-time DMFT would test whether the relevant spectral exponents can be inferred from impurity dynamics before the asymptotic time regime is numerically accessible.

In summary, we have shown that critical real-time dynamics can be reconstructed from
the short time windows accessible to tensor-network simulation by identifying the
branch-cut exponent blindly from the data and imposing it as an algebraic envelope on
an otherwise standard finite-pole fit. A single fixed protocol recovers the known
exponents across the XXZ critical line, at the $z=2$ saturation transition, and in the
gapped free-magnon phase, and it extrapolates severalfold more accurately than
finite-pole methods precisely in the short-window regime where tensor-network data are
expensive. On the noncritical DMFT control the same protocol finds no stable exponent
and safely reduces to the standard pole result. Because the entanglement barrier that
limits real-time simulation is generic, a data-selected treatment of the non-analytic
spectral component can be adopted as a drop-in stage in existing complex-time and
pole-based extrapolation pipelines~\cite{Cao2024,Yu2026,Erpenbeck2026}. More broadly,
blind exponent identification from limited windows addresses a recurring need wherever
anomalous power laws must be extracted from time-limited data --- as, for example, for the
Kardar--Parisi--Zhang superdiffusion observed in Heisenberg spin
chains~\cite{Ljubotina2019,Scheie2021} --- and on experimental and quantum-hardware
platforms whose coherence times bound the accessible window.

\begin{acknowledgments}
This work was supported by the Basic Science Research Program through the National Research Foundation of Korea (NRF) funded by the Ministry of Science and ICT (Grant Nos.~RS-2023-00220471, RS-2025-16064392, RS-2025-23525695, RS-2025-16064392). HSK additionally thanks the KENTECH Startup Funding for new faculty members (202500005A).
\end{acknowledgments}

\appendix
\section{fSD computational details and evaluation process}
\label{sec:protocol}
For all fSD extrapolations we follow the procedure described in Sec.~\ref{sec:esprit}. 
All quantitative comparisons between fSD and the finite-pole baselines are run on the
same fit window $[t_w,t_D]$ and the same subsampled time grid. For fSD, plain ESPRIT,
and LP we additionally use the same root-reflection stabilization (exponentially
growing modes are reflected into the unit disk) and the same data-driven pole count
$M$: the Hankel matrix of the windowed data (pencil parameter $L=N/2$) is factorized
once by singular-value decomposition, and $M$ is the number of singular values above
$10^{-3}$ of the largest. The fSD residual fit is ESPRIT itself, applied to the
detrended signal $t^{\alpha}C(t)$: the $M$-dimensional signal subspace of the same
Hankel matrix yields the residual poles through rotational invariance, and the
amplitudes follow from a least-squares Vandermonde solve. (Replacing this ESPRIT
residual fit by a stabilized linear-prediction fit changes individual short-window
errors of Sec.~\ref{sec:dataeff} by a median factor of $1.2$ but leaves the overall
picture unchanged --- ten of twelve short-window wins for either estimator, with
median advantages of $2.9\times$ versus $3.0\times$ over the full remaining window ---
so the results are not an artifact of the estimator choice.) The Erpenbeck-style baseline instead follows its authors' own
prescriptions --- growing-mode discard and the singular-value threshold $10^{-6}$
(Sec.~\ref{sec:uq}). fSD's single additional degree of freedom, the exponent $\alpha$, is fixed
blind within the fit window and never uses out-of-window data. Concretely, for the XXZ
analysis the fit window starts at $t_w=3.0$, the data are subsampled to a grid spacing
$\delta t=0.25$, the Hankel pencil parameter is $L=N/2$ for ESPRIT ($L=0.4N$ for the
Erpenbeck-style baseline) with the pole count capped at $L-1$, and the blind exponent is
scanned over $\alpha\in[0,1.6]$ on a 41-point grid under the walk-forward validation of
Sec.~\ref{sec:blind}: the fit window is split at the training fractions $0.55$, $0.65$,
and $0.75$, each split is fitted on its training segment alone, and $\alpha^{*}$
minimizes the median of the three validation errors. For a fit cutoff $t_D$ the out-of-window error is
\begin{equation}
\label{eq:oow}
E(t_D)=\frac{\big[\sum_{t_j}\,[\mathrm{Re}\,C_{\rm pred}(t_j)-\mathrm{Re}\,C_{\rm ref}(t_j)]^{2}\big]^{1/2}}
{\big[\sum_{t_j}\,[\mathrm{Re}\,C_{\rm ref}(t_j)]^{2}\big]^{1/2}},
\end{equation}
(where $t_D < t_j < t_{\rm out}$) evaluated on the subsampled grid up to the reflection-free maximum time --- concretely,
the evaluation is cut at $t_{\rm out}=28$ for the $N\ge100$ data sets, $t_{\rm out}=15$
for the $N=50$ field-scan sets, and $t_{\rm out}=27$ for the complex-time $\Delta=0.5$
scan, all safely inside the respective reflection-free windows; all advantage
ratios below are $E_{\rm pole}/E_{\rm fSD}$ on the same grid. In addition to this
full-remaining-window error we quote fixed-horizon errors $E_H$, obtained by
restricting the sums in Eq.~(\ref{eq:oow}) to $t_D<t_j\le t_D+H$ with $H=4$, $8$, and
$16$ (windows for which $t_D+H$ exceeds the reference data are excluded); $E_H$
compares all methods over the same prediction distance for every cutoff, so the
advantage cannot be inflated by the shrinking of the remaining evaluation window at
large $t_D$. For the DMFT retarded
function, which is purely imaginary at the particle-hole-symmetric point studied here,
the same norm is evaluated on the complex modulus $|C_{\rm pred}-C_{\rm ref}|$ instead
of the real part (this also applies to the walk-forward validation of the blind scan).
The norm is global, so zero crossings of the oscillating signal do not destabilize it.
We emphasize the division of roles between the complex signal and its real part: every
fit --- fSD, the baselines, and the inner fits of the blind scan --- is performed on
the full complex signal, while every quoted error and validation loss is evaluated on
$\mathrm{Re}\,C$ as in Eq.~(\ref{eq:oow}). Figures display $|C(t)|$ because the
power-law envelope of the oscillating $\mathrm{Re}\,C$ --- and $|C|$ is precisely that
envelope --- appears as a straight line on logarithmic axes; every deviation curve
shown is nevertheless evaluated on $\mathrm{Re}\,C$, matching the quoted metric.
(Restricting the fits themselves to $\mathrm{Re}\,C$ would discard the phase of the
analytic signal --- each complex mode splits into a conjugate pair, doubling the model
order the same window must support --- and we find that it destabilizes the blind
exponent selection on short windows.)
Short-window medians quoted below take the small-$t_D$ half of each scan's cutoff grid
($t_D\le14$ for the anisotropy scan, $t_D\le12$ for the field scan); the qualitative
conclusions do not depend on this choice. During the blind
$\alpha$ scan, each walk-forward split determines its pole count $M$ from the Hankel
singular values of its own training segment, so neither the validation segments nor
any out-of-window data enter the fit or the order selection; once $\alpha^{*}$ is
chosen, the final extrapolation refits the full window $[t_w,t_D]$ with the window's
own auto-$M$. We also note two structural points. First, the comparison is
data-symmetric but not prior-symmetric: fSD carries one additional data-selected
degree of freedom --- the power-law envelope --- while the pole baselines are
restricted to finite exponential sums. 
Second, root
reflection is a numerical stabilization only: it bounds the extrapolation but does not
enforce causality, positivity, sum rules, or passivity, and such physical constraints
are treated as post hoc diagnostics throughout (in the DMFT spectra of
Appendix~\ref{app:dmftAw}, Fig.~\ref{fig:dmft}, residual negative spectral weight from the broadened transform is
clipped to zero and the sum rule is checked a posteriori). As a reference we also
quote the pole baseline at its \emph{best-case} pole count, defined as the $M$ that
minimizes the true out-of-window error; because this choice uses knowledge of the exact
tail it is favorable to the pole method, and it is reported only to show that the fSD
advantage on short data is not an artifact of the model order.

\section{The rigorous fSD closure: synthetic versus real data}
\label{app:rigorous}

\begin{figure*}[t]
\includegraphics[width=0.80\linewidth]{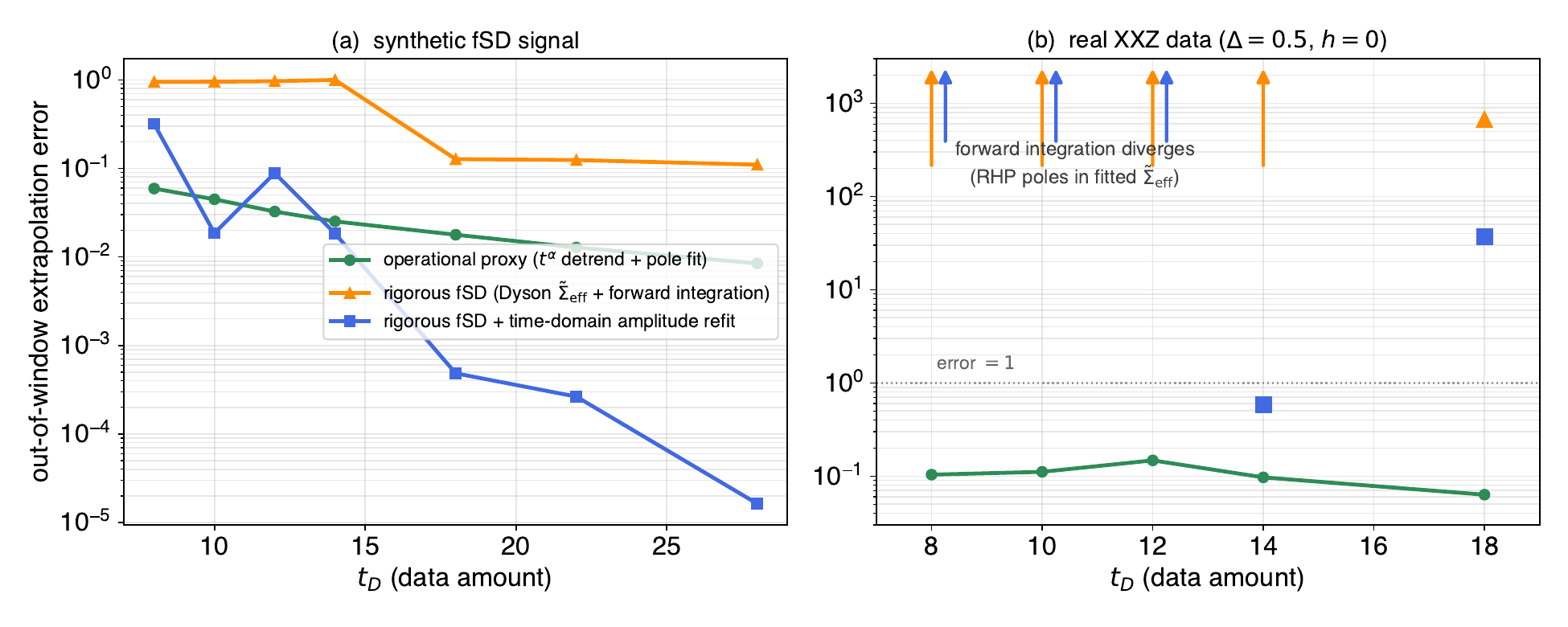}
\caption{\label{fig:rigorous}
Rigorous fSD closure versus the operational proxy. (a) On a synthetic signal
constructed to satisfy the fSD equation ($\alpha=0.5$, two-pole memory kernel), the
rigorous route reaches errors as small as $10^{-5}$ once the kernel amplitudes are
refit in the time domain (blue); without the refit (orange: plain Dyson extraction +
forward integration) it remains above the operational proxy (green) even on synthetic
data. (b) The same three
estimators applied to real XXZ data ($\Delta=0.5$, $h=0$, with $\alpha$ fixed at the
closed-form value): the rational fit of the Dyson-extracted $\tilde\Sigma_{\rm eff}$
acquires right-half-plane poles, and even after these are discarded the truncated
kernel drives the forward integration to divergence for most fit windows (arrows) or to
errors far above unity (isolated symbols), while the proxy remains stable at
$\approx0.1$ (for comparability with the synthetic construction, the appendix proxy
uses a fixed pole count $M=6$ on the raw time grid). This contrast motivates the use of
the proxy for all real-data results.}
\end{figure*}

On signals generated to satisfy Eq.~\eqref{eq:resolvent}, the rigorous route --- Dyson
extraction of $\tilde\Sigma_{\rm eff}(s)$ from the windowed data, a rational fit of its
poles, and forward integration of Eq.~\eqref{eq:fracEOM} --- extrapolates with errors as
small as $10^{-5}$ once the kernel amplitudes are refit in the time domain, orders of
magnitude below the operational proxy on the same windows [Fig.~\ref{fig:rigorous}(a)];
the residual-kernel poles are simultaneously recovered to $3$--$4$ digits (not shown).
Applied to real critical data, the same construction fails [Fig.~\ref{fig:rigorous}(b)]:
the Dyson-extracted $\tilde\Sigma_{\rm eff}(s)$ is no longer exactly rational, its
rational fits acquire right-half-plane (unstable) poles; discarding these still leaves
a truncated kernel whose forward integration diverges outright for most fit windows or
returns errors orders of magnitude above unity, while the proxy remains stable at
$\approx0.1$. This contrast --- exact
closure on data that satisfy the fSD equation by construction, instability on real data
that satisfy it only approximately --- is why all real-data results in the main text
use the proxy of Sec.~\ref{sec:esprit}.

\section{DMFT spectral functions across the Mott transition}
\label{app:dmftAw}

\begin{figure}[t]
\includegraphics[width=0.9\linewidth]{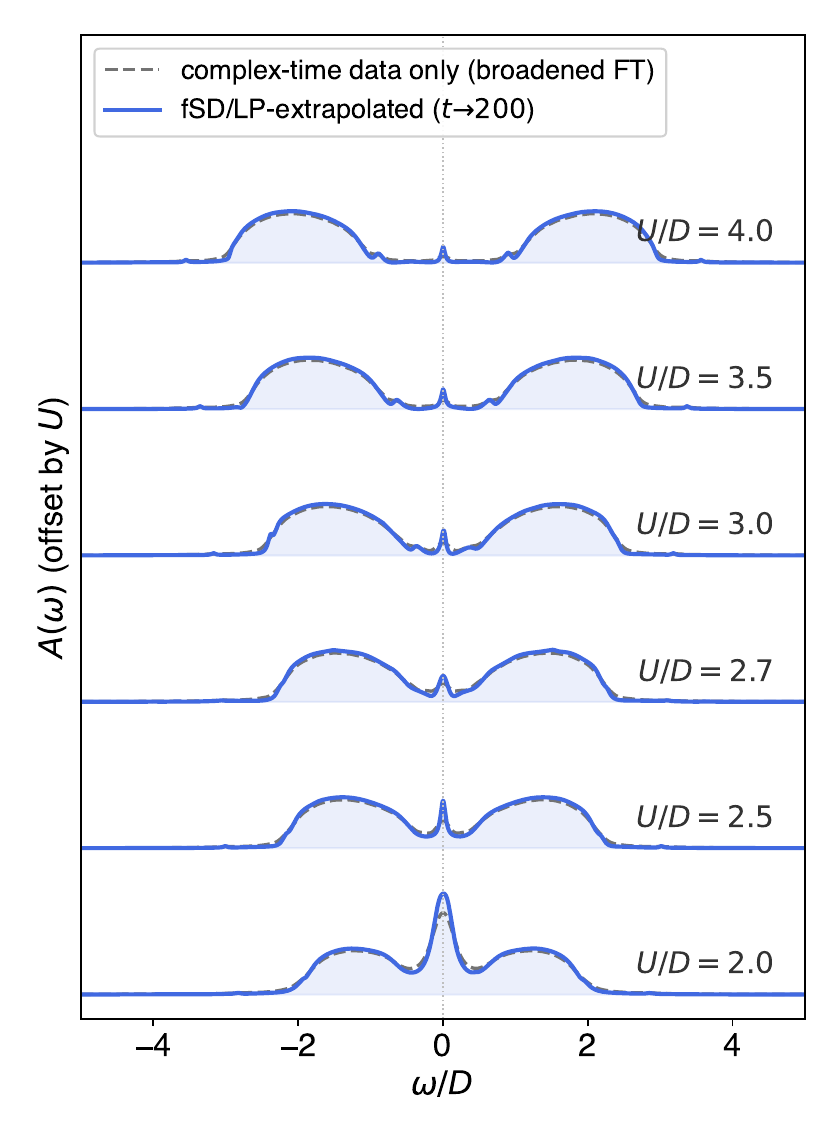}
\caption{\label{fig:dmft}
DMFT spectral function, waterfall over $U/D=2.0\to4.0$ (metal to Mott insulator). Dashed:
$A(\omega)$ from the complex-time data alone (Fourier transform of $G^R(t)$ to $t=30$,
broadening $\eta=0.08D$). Solid (blue): after extrapolating the real-time signal with the
fSD pipeline ($\alpha=0$, reducing to ESPRIT with $M=50$ poles, fit window from $t_w=1.5$) to $t=200$, which permits
the smaller broadening $\eta=0.03D$. The extrapolation resolves the low-energy
quasiparticle peak of the metal [sharp at $U/D=2$, $A(0)\approx2/(\pi D)$] that the
complex-time-only transform smears, and follows its collapse as the Mott gap opens.
(For this noncritical system the blind selection finds no stable fractional order, so
$\alpha=0$ is used and fSD reduces to the pole method.)}
\end{figure}

Figure~\ref{fig:dmft} shows the spectral function $A(\omega)$ of the Bethe-lattice
Hubbard model across the Mott transition ($U/D=2.0\to4.0$) in a waterfall format. The
dashed curves are computed from the complex-time data alone: because the signal ends at
$t=30$, the Fourier transform requires a sizable broadening, which smears the
low-energy structure. The solid curves are obtained after extrapolating the real-time
signal to $t=200$ with the pole-based pipeline; since the blind selection finds no
stable fractional order on these data (Sec.~\ref{sec:falsify}), $\alpha=0$ is used and
fSD reduces to a standard ESPRIT extrapolation. The extrapolation nevertheless adds
clear value: a much smaller broadening becomes admissible, the sharp quasiparticle peak
of the metal [$A(0)\approx2/(\pi D)$ at $U/D=2$] is resolved, and its collapse is
followed as the Mott gap opens. The extrapolated spectra pass the standard diagnostics:
the zeroth moment is $\int A\,d\omega=0.996$ at every $U$, residual negative weight
before clipping stays below $10^{-3}$, and at the metallic point the Fermi-level value
$A(0)=0.621$ agrees with the pinning condition $2/(\pi D)\simeq0.637$ to within
$2.5\%$; these diagnostics are computed by the same script that generates the figure.
Complementary high-resolution real-frequency routes exist for such noncritical spectra,
e.g.\ improved NRG self-energy estimators~\cite{Kugler2022}.

\section{Blind exponent tracking in a field at $\Delta=0.5$}
\label{app:d05}

\begin{figure*}[t]
\includegraphics[width=0.80\linewidth]{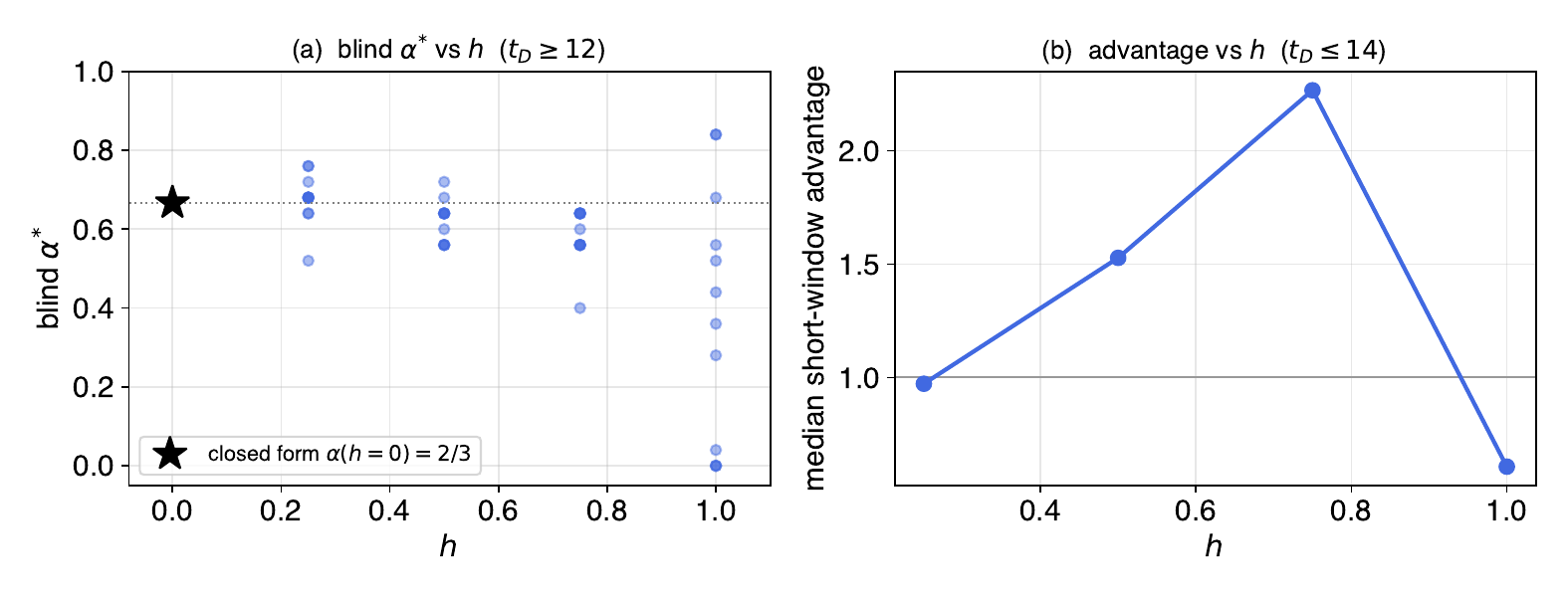}
\caption{\label{fig:d05}
Field scan at $\Delta=0.5$ on complex-time-derived data ($N=100$, eight contours
extrapolated to $\eta\to0$). (a) Blind exponent $\alpha^{*}$ for all $(t_w,t_D)$
windows with $t_D\ge12$ versus $h$; the star marks the closed-form zero-field value
$\alpha=2/3$. The exponent tracks a stable, slowly drifting value up to $h=0.5$ and
destabilizes by $h=1.0$. (b) Median short-window ($t_D\le14$) advantage of fSD over
the best finite-pole baseline.}
\end{figure*}

As a complementary test of the blind selector under a continuously tuned exponent, we
scanned the XXZ chain at $\Delta=0.5$ in a longitudinal field, $h=0.25$--$1.0$
($h_{\rm sat}=1.5$). These data were produced entirely through the complex-time route:
for each $h$, TEBD ($N=100$) evolves a family of eight contours $C(t-i\eta)$ with
$\eta\in[0.25,1.5]$ --- the imaginary shift damps high-energy weight and thereby
suppresses entanglement growth --- and the real-time signal is recovered by polynomial
extrapolation $\eta\to0$ following Ref.~\cite{ChaLeeKim2025}. The exact equal-time
value $C(0)=1/4$ is recovered to within $2.2\%$, and the extrapolation subset-spread is
$\lesssim8\times10^{-3}$ at the start of the analysis window, falling below $10^{-3}$
beyond $t\approx8$--$11$. The blind pipeline of App.~\ref{sec:protocol}
is then applied unchanged.

Figure~\ref{fig:d05} summarizes the outcome. For $h\le0.5$ the blind exponent is stable
and tracks a slowly drifting value --- $\alpha^{*}\approx0.68$ at $h=0.25$ and $0.64$
at $h=0.5$ --- continuously connected to the closed-form zero-field value $2/3$, and
the walk-forward validation loss retains a single sharp minimum. At $h=0.75$ the
short-window selections begin to wander, and by $h=1.0$ the exponent scatters over the
grid with a flat validation landscape: the same multi-component destabilization seen at
$\Delta=1$ below saturation (Sec.~\ref{sec:hscan}), here setting in at roughly two
thirds of the saturation field. The short-window advantage is modest throughout
($0.6$--$2.3\times$; near unity at $h=0.25$ and below it at $h=1.0$). The scan demonstrates two things: the blind selector can track a
field-tuned exponent continuously where a single algebraic component dominates, and its
breakdown provides a practical map of where the single-envelope ansatz stops applying.

\bibliography{refs}

\end{document}